# SPECTROSCOPY OF SUBTHRESHOLD INELASTIC RADIATION-INDUCED PROCESSES IN CRYOCRYSTALS

A.N.Ogurtsov
*Institute of Low Temperature Physics & Engineering of NASU, Kharkov 61103, Ukraine; National Technical University "KhPI", Kharkov 61002, Ukraine*

Abstract: A review is given on the recent spectroscopic studies of the electronic excitation induced processes of large-scale atomic displacements in atomic cryocrystals, including defect formation, atomic and molecular desorption.

Key words: Rare-gas solids, Cryocrystals, Electronic excitations, Self-trapping, Frenkel pairs, Desorption

## 1. INTRODUCTION

Fundamental excitation of non-metallic solids by photons and beams of particles with a kinetic energy below the threshold of knock-on of atoms from the lattice sites – the subthreshold excitation – is a powerful tool for material modification by selective removal of material, controlled changes in selected regions, altering the balance between the process steps, quantum control, etc. [1]. The scission of the bonds, which stabilize the ground-state configuration, by transferring the electronic excitation energy to the lattice requires trapping or self-trapping of electronic excitations [2]. However, the range of materials which exhibit the inelastic subthreshold processes induced by electronic excitation is limited to specific classes of materials, such as alkali halides, alkali earth fluorides and fused quartz [1].

Rare-gas solids (RGS), or atomic cryocrystals, are the model systems in physics and chemistry of solids, and enormous amount of information about electronic excitations in RGS has been documented in several books [2–5] and reviews [6–9]. As a consequence of the closed electronic shells, solid Xe, Kr, Ar, and Ne are the simplest known solids of the smallest binding energy between atoms in the lattice. On the other hand, solid Ar and Ne have band-gap energies, $E_g$, exceeding that of LiF and may be cited as the widest band-gap insulators. Therefore, RGS are very promising systems for study the mechanisms of subthreshold radiation-induced processes.

The present paper reviews the updated understandings of the subthreshold inelastic processes induced by electronic excitations in RGS.

## 2. EXPERIMENTAL

Rare-gas samples exist only at cryogenic temperatures and most of the optical spectroscopy of electronic processes should be done in the vacuum ultraviolet. Making experiments requires an indispensable combination of liquid-helium equipment with windowless VUV-spectroscopic devices and synchrotron radiation as a photon source. To study the electronic excitation energy pathways and a variety of subthreshold inelastic processes, we used the complimentary advantages of cathodoluminescence (possibility to vary the excitation depth beneath the sample surface), photoluminescence (selective-state excitation by synchrotron radiation at high-flux SUPERLUMI-station at HASYLAB, DESY, Hamburg) and thermoactivation spectroscopy (sequential release of electrons from traps of different depth ended in thermoluminescence and thermostimulated exoelectron emission). The experimental setups and methods of sample preparation from vapor phase were described in detail elsewhere [10].

## 3. ELECTRONIC EXCITATIONS AND LUMINESCENCE OF ATOMIC CRYOCRYSTALS

The electronic properties of RGS have been under investigation since seventies [3–7] and now the overall picture of creation and trapping of electronic excitations is basically complete. Because of strong interaction with phonons the excitons and holes in RGS are self-trapped, and a wide range of electronic excitations are created in samples: free excitons (FE), atomic-like (A-STE) and molecular-like self-trapped excitons (M-STE), molecular-like self-trapped holes (STH) and electrons trapped at lattice imperfections. The coexistence of free and trapped excitations and, as a result, the presence of a wide range of luminescence bands in the emission spectra enable one to reveal the energy relaxation channels and to detect the elementary steps in lattice rearrangement.

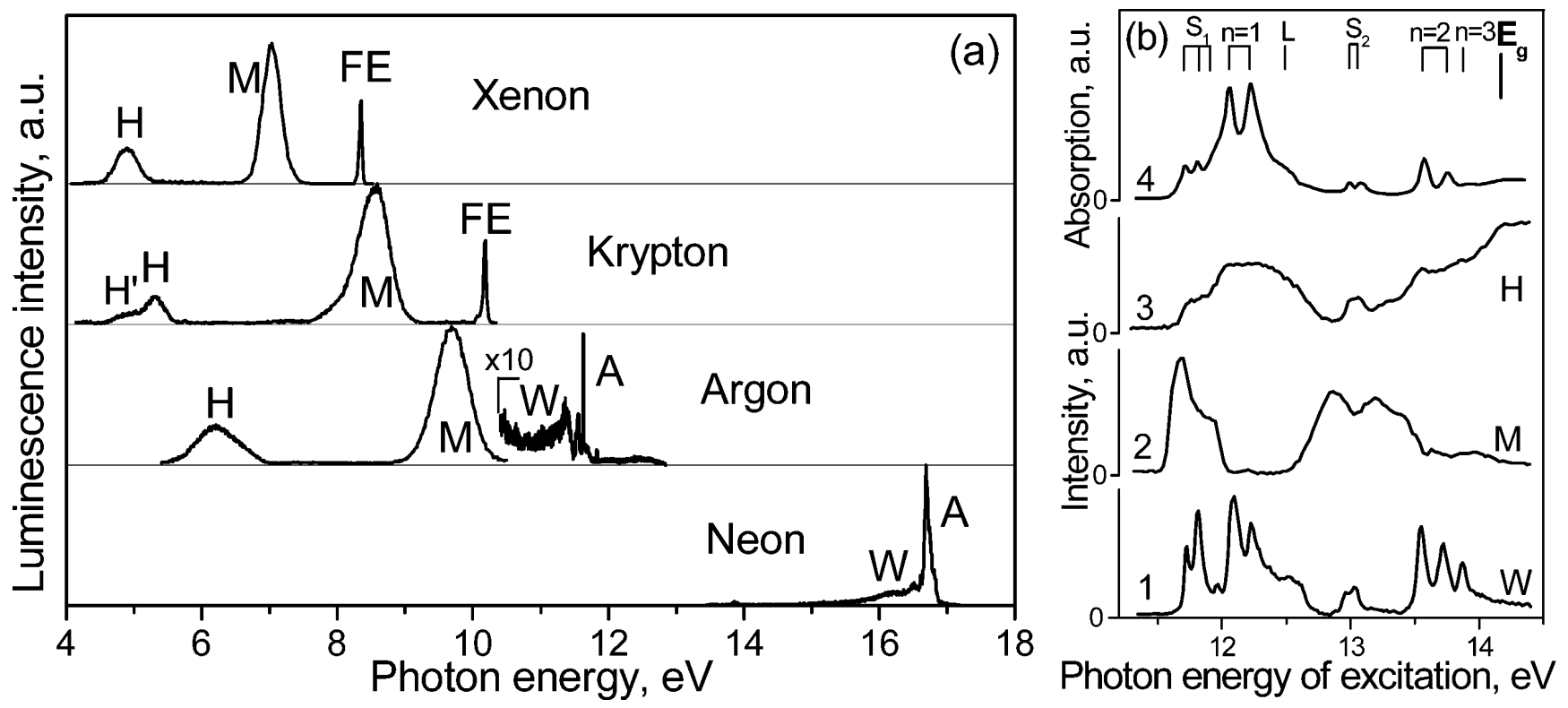

*Figure 1.* (a) – Photoluminescence spectra of atomic cryocrystals at $T$=5 K under excitation energy $h\nu=E_g$. (b) – (1-3) Excitation spectra of molecular luminescence bands of solid Ar; (4) – absorption spectrum of solid Ar (from Ref.7).

The most prominent feature in luminescence of Xe, Kr and Ar – the so-called *M*-band (Fig.1a) – is formed by ${}^{1,3}\Sigma_u^+ \rightarrow {}^1\Sigma_g^+$ transitions in ($R_2$*) excimer M-STE (R=rare gas atom). The negative electron affinity (Table 1) is a moving force of the cavity ("bubble") formation around A-STE in the bulk of crystal, and the desorption of atoms and excimers from the surface of solid Ne and Ar [11]. Radiative "hot" transitions in desorbed excimers of Ar and Ne result in a *W*-band. *A*-bands are emitted by A-STE (R*).

Recent study of charged centers in RGS [12-14] reveal the nature of *H*-bands. These bands correspond to the third continua in rare gas emission [15], which are formed by transitions $(R_2^+)^* \rightarrow (R_2^+)$ in molecular ions [16]. A tiny amount of impurity Xe atoms in solid Kr results in the formation of heteronuclear excited ions $(KrXe^+)^*$ and a corresponding *H'*-band (Fig.1a).

Radiative decay of free excitons from the bottom of the lowest Γ(3/2), $n$=1 excitonic band produces strong lines *FE* (Fig.1a) in spectra of solid Xe,

*Table 1.* Some parameters of energetic structure and luminescence of RGS [2,7]

| | Xe | Kr | Ar | Ne |
|---|---|---|---|---|
| Band-gap energy at the Γ-point, $E_g$ (eV) | 9.298 | 11.59 | 14.14 | 21.58 |
| Binding energy per atom, $\varepsilon_b$ (eV) | 0.172 | 0.123 | 0.088 | 0.026 |
| Electron affinity, $\chi$ (eV) | 0.5 | 0.3 | −0.4 | −1.3 |
| Bottom of the lowest excitonic band Γ(3/2), *FE* (eV) | 8.37 | 10.14 | 12.06 | 17.36 |
| Barrier height against M-STE formation, $H^M_{max}$ (meV) | 20 | 10 | 2 | 0.3 |
| Barrier height against A-STE formation, $H^A_{max}$ (meV) | | 30 | 10 | 1 |
| Energy release at M-STE formation, $E_M$* (eV) [17] | 0.45 | 0.75 | 1.25 | |
| Energy release at A-STE formation, $E_A$* (eV) [17] | | | 0.2 | 0.58 |
| Threshold energy for $M_1$ subband, $E_1$ (eV) [18] | 8.18 | 9.87 | 11.61 | |
| Threshold energy for $M_2$ subband, $E_2$ (eV) [18] | 8.28 | 10.03 | 11.81 | |

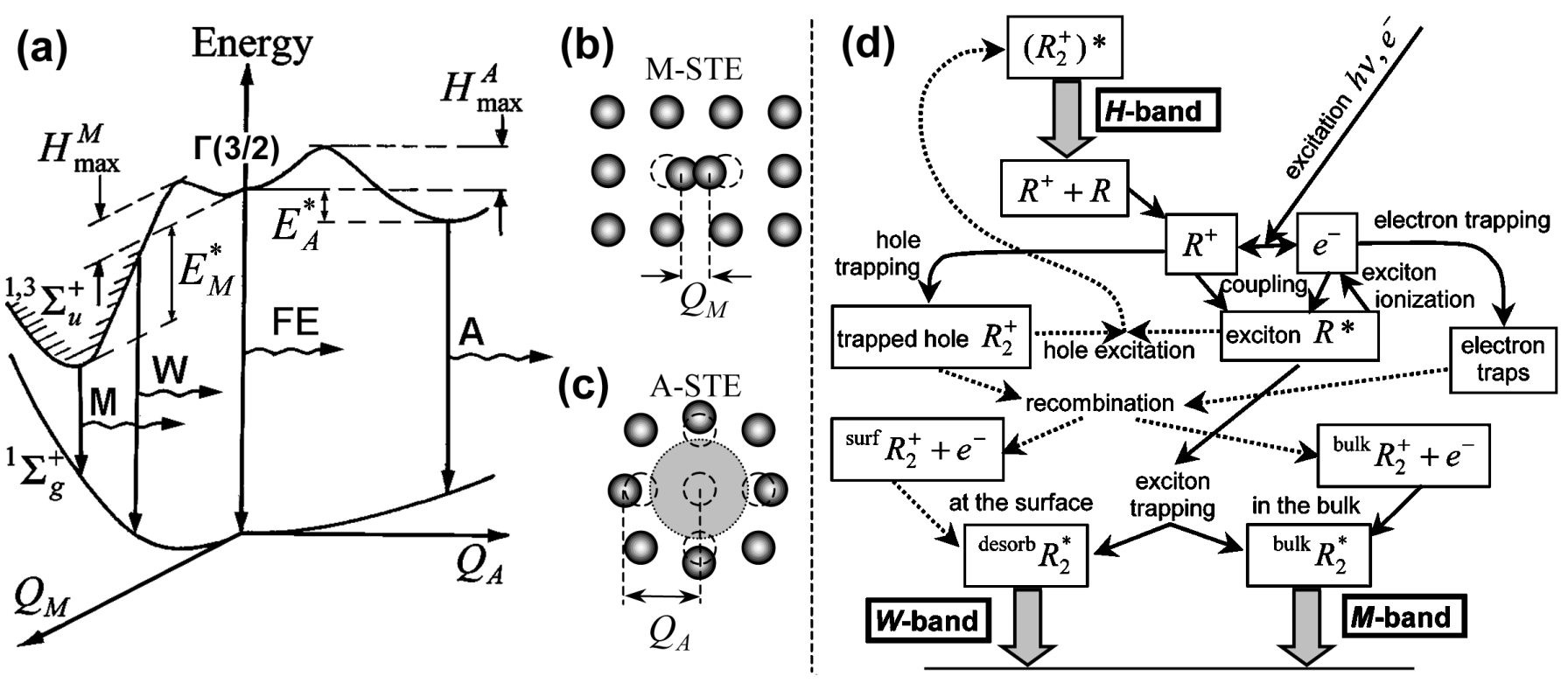

*Figure 2.* (a) – Potential curves of M-STE and A-STE. (b,c) – Simplified geometrical structure of M-STE and A-STE. (c) – Scheme of the population of molecular neutral and charged trapped centers in atomic cryocrystals.

Kr and Ar [19].

The schematic representation of A-STE and M-STE and luminescence transitions in neutral centers is shown in Fig.2a. In M-STE the configuration co-ordinate, $Q_M$, is an internuclear distance of the molecule (Fig.2b). In A-STE (Fig.2c), the configuration co-ordinate, $Q_A$, is a radius of microcavity (nearest-neighbor distance) [7].

The commonly used scheme of energy relaxation in RGS includes some stages (Fig.2d, solid arrows). Primary excitation by VUV photons or low energy electrons creates electron-hole pairs. Secondary electrons are scattered inelastically and create free excitons, which are self-trapped into atomic or molecular type centers due to strong exciton-phonon interaction.

Complementary studies of neutral [20] and charged [16] intrinsic trapped centers, comparison of cathodoluminescence [21] and thermoluminescence data [12] with results of analysis of photoelectron scattering [13] and pump-probe experiments [14] allow us to extend the energy relaxation scheme (Fig.2d, dotted arrows) including electron-hole recombination channels. The formation of *H*-band emitting centers $(R_2^+)^*$ occurs through the excitation of STH by an exciton. The bulk recombination of trapped holes with electrons populates the $(R_2^*)$ states with subsequent *M*-band emission [22]. After surface recombination of STH with electrons the excited dimers escape from the surface of the crystal with subsequent *W*-band emission.

The general behavior of the excitation spectra of *H*- and *W*-bands (Fig.1b) reproduces all surface (*S*) and bulk (transverse *n*=1,2,3 and longitudinal *L*) features of photoabsorption. The opposite behavior of the *M*-band underlines the branching of competing channels of population neutral and charged trapped centers [16].

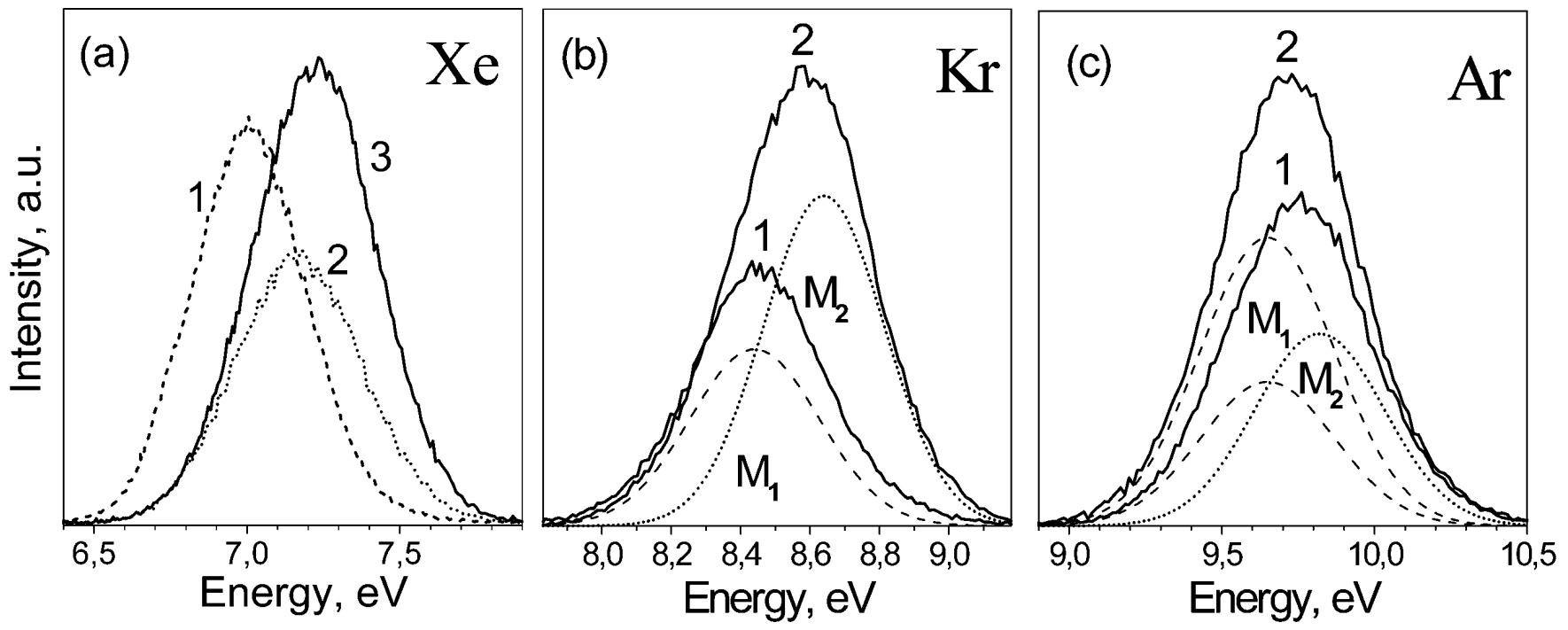

*Figure 3. M*-bands of RGS at *T*=5 K under photoexcitation with energies listed below. (a) Xe – (1) 8.25 eV, and (2,3) 8.89 eV before (2) and after (3) annealing at 60 K. (b) Kr – (1) 10.02 eV, and (2) 10.6 eV. (c) Ar – 12.06 eV before (1) and after (2) 15 min of wide-band synchrotron-beam irradiation with energies centered at $h\nu$=20 eV with FWHM=13 eV.

# 4. FRENKEL PAIRS FORMATION BY ELECTRONIC EXCITATIONS

The interatomic bond scission in the crystal lattice may be simulated either by elastic encounters between atoms composing solids and incoming particles or by creation of electronic excitations which transfer the energy to a specified crystal cell. The energy stored by electronic excitations in RGS is much higher than the binding energy $\varepsilon_b$ (Table 1), and various trapping processes concentrate the energy within a volume of about a unit cell. The extremely high quantum yield of luminescence [23] allows one to neglect non-radiative transitions, and the population of antibonding $^1\Sigma_g^+$ ground molecular state is usually considered as a main source of kinetic energy for a large-scale movement of atoms finishing in the Frenkel defects or desorption of atoms in the ground state – ground-state (GS) mechanism.

On the other hand, the processes of formation of A-STE and M-STE centers themselves are accompanied by a considerable energy release ($E_A$* and $E_M$*) to the crystal lattice (see Fig.2a. and Table 1) which also exceeds the binding energy $\varepsilon_b$. Such an excited-state (ES) mechanism of the large-scale atomic movement was the subject of our recent investigations.

The analysis of the luminescence spectra of solid Xe, Kr, and Ar (Fig.3) under different excitation conditions, excitation energies and crystal-growth conditions made it possible to elucidate the internal structure of *M*-bands. Each of the *M*-bands can be well approximated by two Gaussians: low energy subband $M_1$ and high energy one $M_2$. The subband $M_2$ is dominant in the luminescence of more perfect samples. The spectra of samples with a great number of initial defects are mainly determined by the component $M_1$.

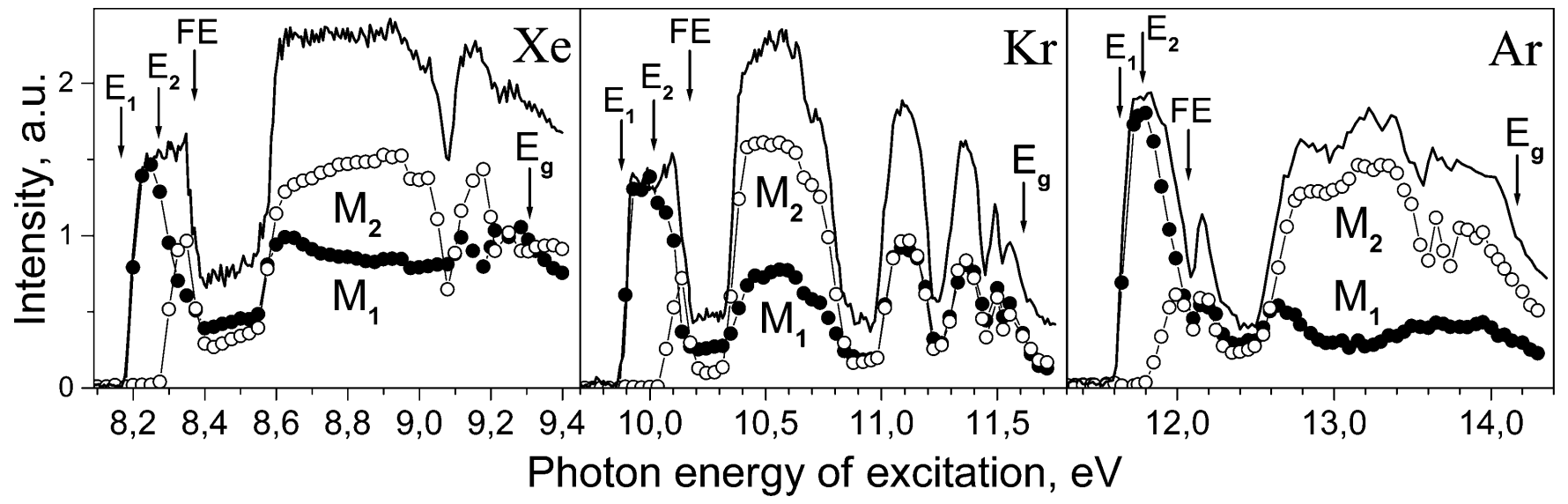

*Figure 4*. Excitation spectra of *M*-bands and subbands $M_1$ (●) and $M_2$ (○) of RGS.

This suggests that the subband $M_2$ is emitted by the excitons which are self-trapped in the regular lattice ($M_2$-centers) while the component $M_1$ is emitted by the centers ($M_1$-centers) which are populated during trapping that occurs with the lattice imperfections involved.

Because of the strong overlapping of molecular subbands there is no way of recording the excitation spectra of the $M_1$ and $M_2$ subbands separately and we restored these spectra by decomposing the sequence of the luminescence spectra measured at different excitation energies (Fig.4) [18]. Using the restored excitation spectra one can determine their threshold energies $E_1$ and $E_2$ for the $M_1$ and $M_2$ subbands to appear (Table 1). In all cases the excitation spectra of the $M_1$-subbands exhibit preferential excitation at energies below the $n=1$ exciton and in the range $E_1<E<E_2$ the direct photoabsorption by defect-related centers produces the only $M_1$ component in the spectra [24]. The transformation of the *M*-band due to the sample annealing (Fig.3a) or irradiation, resulting in the lattice degradation (Fig.3c), in all cases may be described by the intensity redistribution between two subbands.

The *A*-bands of solid Ar and Ne have a similar two-component internal structure [6,19]. Each band of the bulk emission associated with the transitions $^3P_1 \rightarrow {}^1S_0$ and $^1P_1 \rightarrow {}^1S_0$ consists of a high-energy component stemmed from A-STE in a regular lattice and a low-energy one which appears to be associated with structural defects.

The time (or dose) dependences of the subbands turned out to be a very precise and sensitive tool to study the defect formation, and STE can be used to both trigger and probe the dynamics of lattice rearrangement in RGS. Figures 5 and 6 show examples of evolution of luminescence spectra of RGS under irradiation by electrons (Fig.5) and photons (Fig.6). In all cases a pronounced increase in the intensity of the defect component during irradiation has been observed for A-STE (Fig.6a–c), M-STE (Fig.5a) and for impurity atomic emissions (Fig.5b) in solid Xe [10,18,19], Kr [18,20], Ar [18,22,25] and Ne [6,25,26], evidencing the accumulation of stable long-lived defects in the lattice. Since the energy of STE is transferred into kinetic energy of atomic motion over a unit cell, the formation of three-, two-, or one-dimensional defects is ruled out. In this case, only the point defects, Frenkel-pairs, may emerge in the bulk of the crystal.

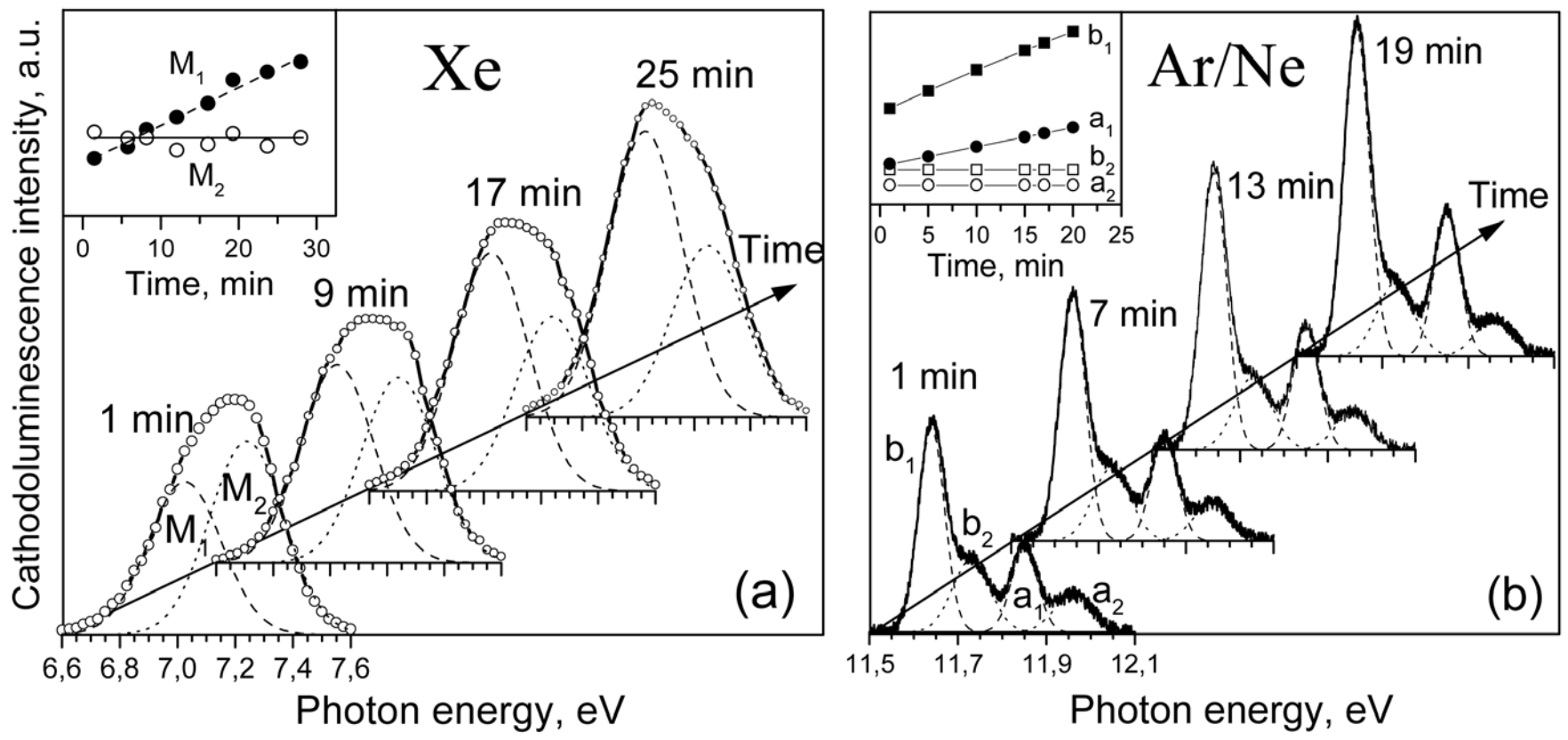

*Figure 5.* Evolution with exposure time (dose) of cathodoluminescence of solid Xe (a) and Ar atoms in Ne matrix (b). The insets show the dose dependences of the subbands.

M-STE is the main channel of exciton self-trapping in solid Xe, Kr, and Ar. In addition to the GS-mechanism, we proposed an ES-mechanism of M-STE to Frankel-pair conversion [20] which consists of three stages (Fig.7). The process is supposed to occur by (stage 1) self-trapping of an exciton (Fig.7a→b) with a subsequent displacement (stage 2) of M-STE from the centrosymmetric position in the <110> direction (Fig.7b→c) followed by (stage 3) reorientation to the <100> direction (Fig.7d) to stabilize the defect. The radiative decay of the stabilized excimer (Fig.7d) results in the creation of a stable point defect, Frenkel-pair, in the lattice, whereas the radiative decay of M-STE in the off-center position (Fig.7c) returns the lattice into the initial (Fig.7a) state without permanent defect. Thus, the state Fig.7c may be considered as a metastable short-lived lattice defect, which, together with stable defects, emits the $M_1$-subband, but is not accumulated in the crystal lattice. The $M_2$-subband is emitted by radiative decay of M-STE in the on-center position Fig.7b.

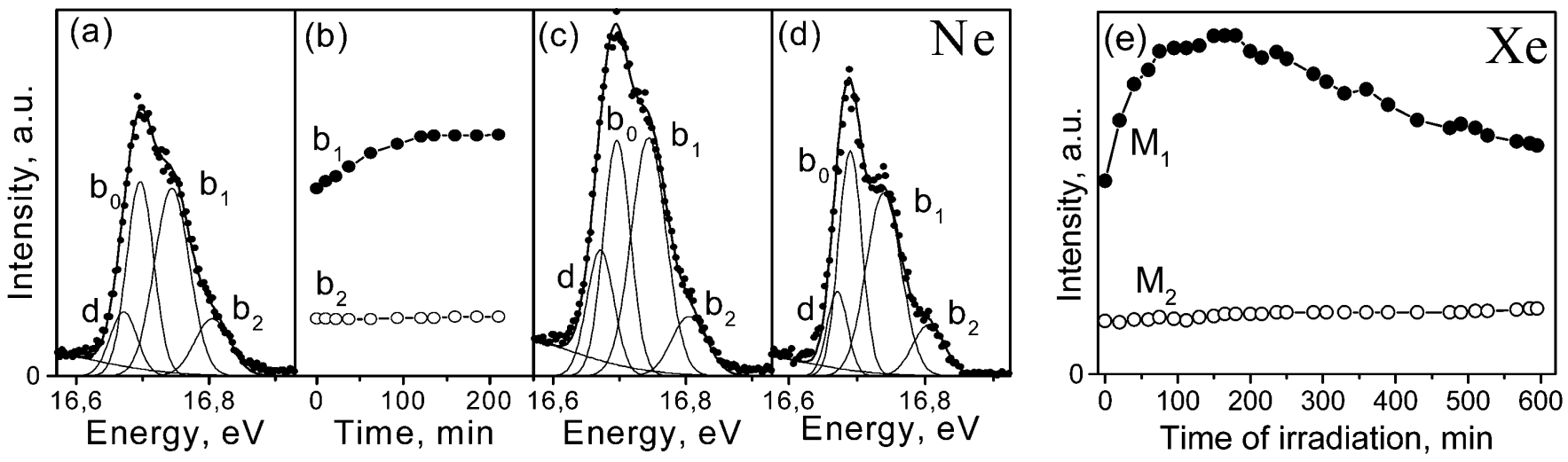

*Figure 6.* (a)–(d) Evolution of *A*-band of solid Ne at T=5 K: (a–c) – under irradiation by 20 eV photons, (d) – after annealing at T=11.5 K. (e) – Dose dependence of $M_1$ and $M_2$ subbands of solid Xe excited by photons with $h\nu$=9.15 eV at $T$=21 K.

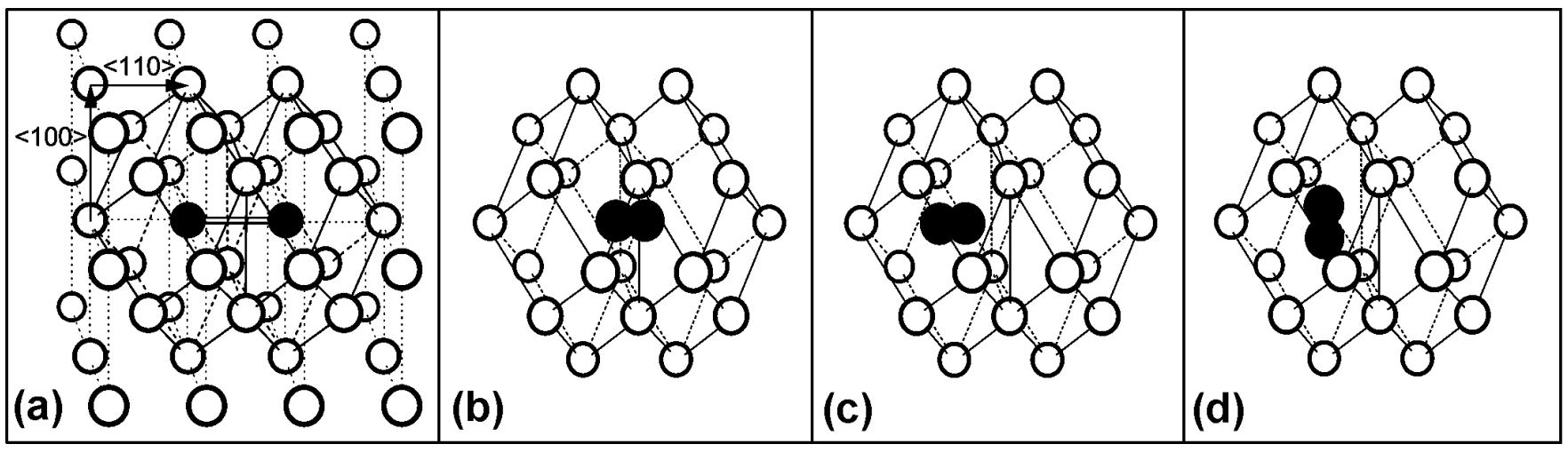

*Figure 7.* Scheme of ES-mechanism of Frenkel-pair formation induced by exciton self-trapping into quasi-molecular state.

Increase of the stable Frenkel-pair concentration under irradiation of the samples is saturated (Fig.6) when the trapping of excitons at defects exceeds the exciton self-trapping in the perfect lattice. Further long-time irradiation of the samples results in an aggregation of vacancies and interstitials, which results in decrease of intensity of defect subbands (Fig.6e).

The Frenkel-pair formation induced by excitation of Rydberg states of atomic-like centers was studied both for the intrinsic process of lattice degradation (exciton self-trapping in solid Ne [26]) and for the extrinsic process of lattice degradation induced by excitation of impurity atoms (trapping of exciton at Ar impurity in Ne matrix [25] (Fig.5b) and selective photoexcitation of Xe impurity in Ar matrix [27]). The strong repulsion of the Rydberg electron with a closed shell of surrounding atoms induces a substantial local lattice rearrangement. In solid Ar and Ne this rearrangement leads to a "bubble" formation, i.e., the surrounding atoms are pushed outwards and a cavity is formed around the excited atom (Fig.2c and Fig.8a).

The ES-mechanism of Frenkel-pair formation as a result of excitation of Rydberg atomic states was confirmed by recent molecular dynamics calculations [28,29]. After the bubble formation the surrounding ground state atoms appear to have moved to the second shell. It was found that the second-nearest neighboring vacancy-interstitial pairs could create the permanent defects, which remain in the lattice after exciton annihilation (Fig.8b) [29].

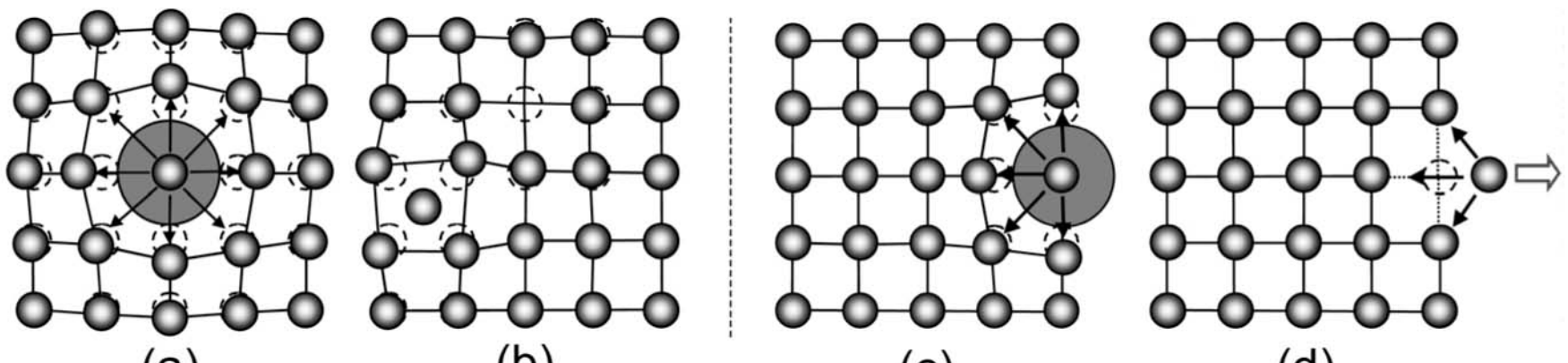

*Figure 8.* Scheme of inelastic processes induced by A-STE in atomic cryocrystals: (a→b) – Frenkel-pair formation; (c→d) – desorption of excited atoms.

In all RGS selective excitation of excitons by photons of energies below the band-gap energy $E_g$ results in accumulation of Frenkel-pairs, which is a direct proof of the excitonic nature of the ES-mechanism of defect formation.

## 5. DESORPTION INDUCED BY ELECTRONIC EXCITATIONS

The ejection of atoms or molecules from the surface of solid in response to primary electronic excitation is referred to as electronically stimulated desorption (ESD) or desorption induced by electronic transitions (DIET). Localization of electronic excitations at the surface of RGS induces DIET of atoms both in excited and in ground states, excimers and ions. Most authors (see e.g. Refs. [8,11,23,30] and references therein) discuss their results on DIET from RGS in terms of three different desorption mechanisms namely (i) M-STE-induced desorption of ground-state atoms; (ii) "cavity-ejection" (CE) mechanism of desorption of excited atoms and excimers induced by exciton self-trapping at surface; and (iii) "dissociative recombination" (DR) mechanism of desorption of excimers induced by dissociative recombination of trapped holes with electrons.

(i) After the radiative decay of M-STE (Fig.2a) the strongly repulsive $^1\Sigma_g^+$ ground molecular state is populated. The nonradiative transfer of electronic energy to the kinetic energy of two involved atoms induces the Frenkel-pair formation (GS-mechanism) in the bulk of the sample or ESD of the ground-state atom, if such a process occurs at the surface. Desorption occurs either directly by the escape of one of these "fast" atoms or through secondary processes like collision cascades initiated by the fast particles.

(ii) The "cavity-ejection" mechanism (Fig.8c,d) is a consequence of repulsive force between the excited atom (A-STE) and its neighbors due to the overlap of their electron wave functions. For solid Ne and Ar which have a negative electron affinity (Table 1) the free exciton self-trapping in the bulk results in a cavity ("bubble") formation around A-STE (Fig.8a). At the surface the short-range repulsion is no longer spherically symmetric like in the bulk but effectively directed outside the surface, leading to the ejection of the excited atom (Fig.8d) or excimer [11]. DIET by the CE-mechanism has unambiguously been verified by the detection of luminescence light originated from electronically excited species in the bloom of gas in front of the surface under irradiation [11,30], and by the detection of metastable particles in ESD [31]. In luminescence spectra the desorption of excited atoms manifests itself as narrow resolution-limited atomic lines [10,19] (e.g. line $b_0$ of solid Ne in Fig.6a-d).

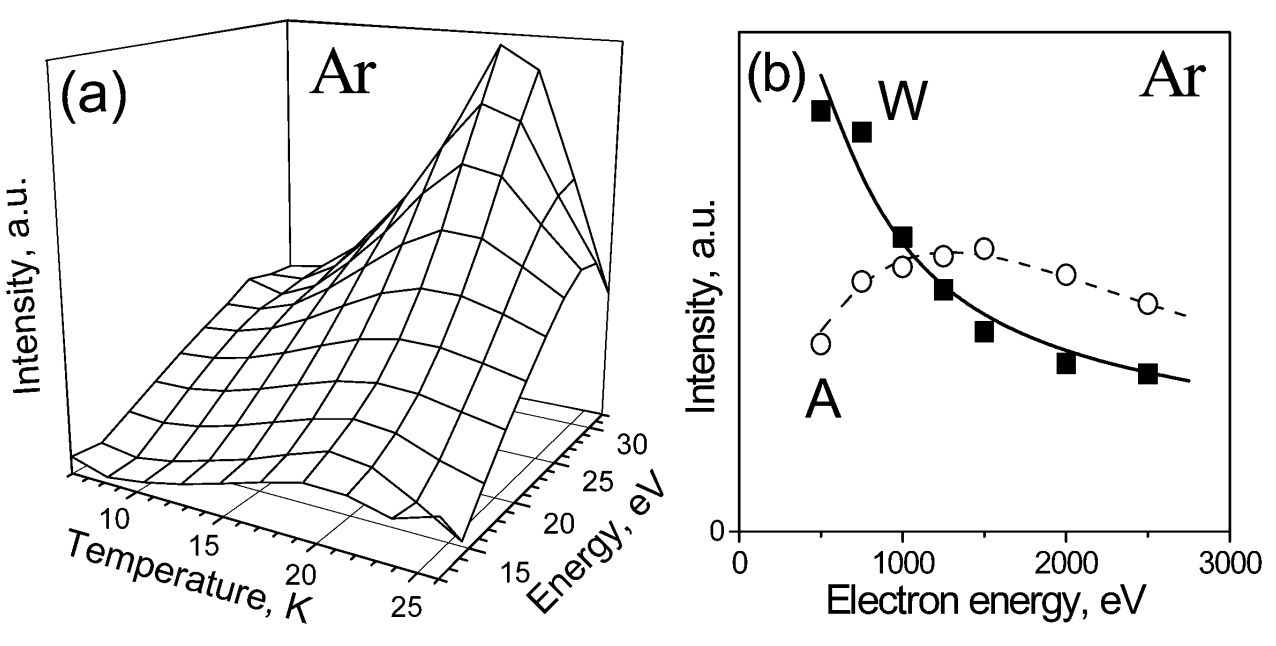

*Figure 9.* (a) – The dependence of integral intensity of *W*-band of solid Ar on temperature and photoexcitation energy. (b) – The dependence of integral intensity of *W* and *A* bands on energy of electrons in cathodoluminescence of solid Ar.

(iii) The DR-mechanism was proposed to explain the desorption of excimers from the surface of solid Ar. In luminescence spectra the DIET of vibrationally excited dimers of Ar manifests itself as a *W*-band (Fig.1a). The recombination of self-trapped hole $Ar_2^+$ with an electron at the surface of the sample results in the separation of the Ar and Ar* atoms along the repulsive $3\Pi_u$ potential of $Ar_2^*$. The dimerization of the moving energetic Ar* with a neighboring Ar results in the desorption of $Ar_2^*$ [11].

Direct verification of DR-mechanism of DIET was provided [21] by combining the state-selective photoexcitation of the sample and the controlled thermally induced release of electrons from electron traps (Fig.9a). In RGS, after electron-hole pair creation at selective excitation by photons with energies $E>E_g$, the hole may survive and be self-trapped if the electron is captured by any kind of traps [32]. In solid Ar at $T>21$ K the main part of electron traps is not active [12], the electron-hole recombination occurs before self-trapping the holes, and, therefore, the concentration of *W*-band emitting centers decreases (Fig.9a). On the contrary, the heating above T=21 K at selective excitation of solid Ar in the excitonic energy range $E<E_g$ produces no decrease in the intensity of the *W*-band.

Cathodoluminescence provides an additional evidence of non-excitonic nature of the mechanism of excimer desorption. The dependences of integral intensity of luminescence bands on electron beam energy (Fig.9b) for *A*-band (excitonic CE-mechanism) and *W*-band (non-excitonic DR-mechanism) are quite different. Changing the electron energy varies the excitation depth beneath the sample surface. The electronic excitations propagate through the bulk as free excitons or mobile holes and are localized at the surface, causing desorption distant from the site of primary excitation [8]. The mobility of holes is much smaller than that of free excitons [33], and just the energy transfer to the surface by excitons or holes governs the efficiency of DIET of excited atoms (*A*-band) or excimers (*W*-band).

## 6. CONCLUSIONS

Atomic cryocrystals which are widely used as inert matrices in the matrix isolated spectroscopy become non-inert after excitation of an electronic subsystem. Local elastic and inelastic lattice deformation around trapped electronic excitations, population of antibonding electronic states during relaxation of the molecular-like centers, and excitation of the Rydberg states of guest species are the moving force of Frenkel-pairs formation in the bulk and desorption of atoms and molecules from the surface of the condensed rare gases. Even a tiny probability of exciton or electron-hole pair creation in the multiphoton processes under, e.g., laser irradiation has to be taken into account as it may considerably alter the energy relaxation pathways.

## ACKNOWLEDGEMENTS

I have the pleasure of expressing my thanks to Prof. G. Zimmerer and Dr. E. Savchenko for fruitful discussions. The support of DFG grant 436 UKR 113/55/0 is gratefully acknowledged.